\documentclass[showpacs,aps,prb,twocolumn]{revtex4}
\usepackage{graphicx}
\usepackage{amsmath}
\usepackage{amssymb}

\begin{document}

\bibliographystyle{apsrev}
\title{Strong 3D correlations in vortex system of Bi2212:Pb}

\author{L.~S.~Uspenskaya and A.~B.~Kulakov}
\affiliation{Institute of Solid State Physics, Russian Academy of
Sciences, Chernogolovka, Moscow Distr., 142432, Russia, e-mail:uspenska@issp.ac.ru}

\author{A.~L.~Rakhmanov}
\affiliation{Institute for Theoretical and Applied
Electrodynamics, Russian Academy of Sciences, Izhorskaya Str.
13/19, Moscow, 125412 Russia, e-mail: andreyr@orc.ru}
\date{\today}

\begin{abstract}
The experimental study of magnetic flux penetration under crossed 
magnetic fields in  Bi2212:Pb single crystal performed by magnetooptic 
technique (MO) reveals remarkable field penetration pattern alteration 
(flux configuration change) and superconducting 
current anisotropy enhancement by the in-plane field. The anisotropy increases 
with the temperature rise up to $T_m = 54 \pm 2 K$. At $T = T_m$ an abrupt change 
in the flux behavior is found; the correlation between the in-plane 
magnetic field and the out-of-plane magnetic flux penetration disappears. 
No correlation is observed for $T > T_m$.
The transition temperature $T_m$ does not depend on the magnetic field strength. 
The observed flux penetration anisotropy is considered as an evidence of a strong 3D %%@
correlation between pancake vortices in different CuO planes at $T < T_m$.     
This enables understanding of a remarkable pinning observed in Bi2212:Pb
at low temperatures.
\end{abstract}

\smallskip

\maketitle

\section{Introduction} \label{In}

The structure and dynamics of the magnetic flux in high-$T_c$
superconductors (HTSC) are intensively studied because of their
importance for both a fundamental physics and applications. The
main results of these studies are summarized in several 
reviews.~\cite{Brandt,Fein,Blat} The investigations reveal the variety of
flux line lattice (FLL) states or phases in HTSC. 
Peculiarities of the FLL structure are determined by the crystal
symmetry or by the origin of pinning. The transitions between 
different FLL phases are possible with
temperature and magnetic field variation. The qualitative
difference in the FLL properties is found to be closely related 
to the layered
structure of crystal lattice of low and high
anisotropic HTSC materials. In particular, the pancake-like
and Josephson-like vortex structures are observed in inclined
magnetic field in highly anisotropic Bi- and Tl-based  superconductors,
whereas anisotropic Abrikosov vortices are found in superconductors with 
a lower anisotropy, such as YBCO.
However, even in the highly anisotropic superconductors
the three-dimensional (3D) correlations can exist between
two-dimensional the (2D) pancake-like vortices located in neighboring
CuO planes.~\cite{Brandt,Fein,Blat} 
The FLL behavior of such 3D-correlated phase is in some
aspects closer to the anisotropic Abrikosov vortex lattice than to
uncorrelated 2D phase. Basically, the 3D correlations
in the pancake structure could disappear with the increase of
temperature due to the first-order phase transition or FLL
melting. A possibility of different types of the FLL phase
transitions in highly anisotropic HTSC was widely %%@
discussed.~\cite{Brandt,Fein,Blat,melt,kosh,baz,dod,Zeld,InvM,ir1,ir2,ir3,ir4}

The high resolution magnetooptic (MO)
technique~\cite{revMO,revMO1} is admitted to be a convenient tool for direct
observation of the magnetic flux structure and dynamics in
superconductors. In particular, the MO studies in crossed magnetic
fields are employed to clarify a presence or absence of the 3D
correlations in FLL of superconductor single
crystals.~\cite{In,Zeld,InvM,Vl-Vl,Grig,Tokun1,Tokun2,Mats}  
Usually, a plate like sample of a single crystal is placed in these 
experiments in a DC magnetic field directed in the 
sample plane, $\mathbf{H}_{ab}$, and then a field, $\mathbf{H}_{z}$,
perpendicular to the plane  is applied. In such geometry the MO
technique is used to study a penetration of the magnetic flux
induced by the field $\mathbf{H}_{z}$. The experiments reveal
two strikingly different types of flux behavior depending on
the anisotropy of the material.~\cite{In} The transverse flux
moves into the YBCO single crystals preferably along the
direction of the in-plane magnetic field $\mathbf{H}_{ab}$. Quite
contrary, the transverse magnetic flux penetrates 
independent of the orientation of the in-plane magnetic field in
case of highly anisotropic Bi2212 superconductors.
This difference can be readily understood.~\cite{In,kes,br} 

Two systems of orthogonal vortices evidently exist in the 
sample under considered field configuration. First system is 
induced by the in-plane field and the second one enters the 
crystal under the growing perpendicular 
field. In case of moderate crystal 
anisotropy, both systems are the systems of mutually perpendicular
Abrikosov vortices. It is rather evident that perpendicular flux
lines moves easier along the in-plane vortices than across them, 
because vortex intersection requires  additional driving 
force and additional energy.~\cite{cross}
This mechanism is effective for Abrikosov-like vortices and
insignificant for the pancake-like structures existed in materials
with high anisotropy. 

The second reason for the in-plane field
induced anisotropy is related to the magnetic interaction between
vortices. This interaction gives rise to the so-called
force-free configuration~\cite{CampIv} at which vortices are twisted 
around the in-plane field. The critical current density, $\mathbf j_{cf}$, 
is limited in this case by the specific FLL instabilities.~\cite{CampIv,ff1,ff2,ff3,ff4} 
This current is usually remarkably higher than the current 
determined by pinning, $\mathbf j_{cf}\gg \mathbf j_{cp}$. 
This leads to the situation when that the current screening 
the flux motion along the $\mathbf{H}_{ab}$ is much smaller than
the current screening the motion across $\mathbf{H}_{ab}$. 
This mechanism is important for Abrikosov vortices. 
It could be effective also for the pancake structure in case of strong 
enough correlation between the pancakes located
in different CuO planes. Obviously, an interaction between Josephson 
vortices and the pancakes should be significantly weaker.~\cite{In,kes}

However, the absence of strong anisotropy of the magnetic flux
penetration induced by the in-plane field,~\cite{In} does not mean that
the pancakes and Josephson vortices are completely independent in
Bi2212. A specific weak interaction is discussed widely and 
acknowledged by direct MO observation under low
magnetic fields in a wide temperature 
range.~\cite{Zeld,InvM,Vl-Vl,Grig,Tokun1,Tokun2,Mats,bul,sav}

The transition between two mentioned above types of magnetization 
behavior of superconductors is not reported in
the literature. Probably such type of transition could be observed in
the material with an intermediate anisotropy compared to YBCO and
Bi2212. Possible way to achieve the goal is to increase the
anisotropy of the YBCO material by preparing of oxygen-deficient
samples or to reduce the anisotropy of the Bi2212 superconductor by Pb
doping. 

It is well known, that Pb doping in Bi2212 single
crystals~\cite{baz,Pb1,Pb2,Pb4,Pb} 
reduces the electromagnetic anisotropy parameter $\gamma^2=\rho_c/\sqrt{\rho_a \rho_b}$
from $8.5\cdot10^3$ downto $2.5\cdot10^3$,  where
$\rho_i$ denotes the normal resistivity along the corresponding
crystal axis $\mathbf{\emph{i}}$, as measured at $T=100$~K
 with Pb content varying from 0 to 0.3. Besides, the doping significantly
increases the critical current density that was attributed to the pinning
at so-called 'laminar' superstructure formed by
variation of the Pb concentration in a system of planes
parallel to the $\mathbf{ac}$ crystal plane.~\cite{Pb}

In the present work a penetration of the transverse magnetic
flux into the Pb-doped Bi2212 single crystals magnetized by the
in-plane magnetic field is studied by the MO technique in a wide 
temperature range, $12\div 91$~K. We characterize the flux penetration 
qualitatively by images and quantitatively by the profiles of the perpendicular
magnetic induction $B_z(\mathbf{r})$ measured along different 
directions in the sample plane. We find that the in-plane magnetic 
field $\mathbf{H}_{ab}$ influence remarkably on the transverse 
flux penetration pattern if the temperature $T$ is 
lower than some threshold value $T_m = 54\pm 2$~K. 
The $\mathbf{H}_{ab}$ increases the current anisotropy and 
causes the preferential flux propagation along its direction. 
The current along the $\mathbf{H}_{ab}$ 
becomes stronger while the current across the $\mathbf{H}_{ab}$ becomes
weaker. Both the current and flux penetration anisotropy 
 increase with the $\mathbf{H}_{ab}$ strength and 
temperature rise.

The flux behavior changes drastically at $T = T_m$. In the temperature 
range $T\geq T_m$ the penetration becomes 
independent of the in-plane field direction and the flux creep increases
significantly. The transition temperature $T_m$ is independent of
the strength of both magnetic fields $\mathbf H_{ab}$ and $\mathbf H_z$ 
within the studied range $0<\mathbf H_{ab} < 1800$~Oe and
$0<\mathbf H_z < 300$~Oe. So, our samples behave like YBCO at temperatures
below $T_m$ and like Bi2212 at higher temperatures. 

\section{Samples} \label{Samp}

The single crystals of
(Bi$_{0.7}$Pb$_{0.3}$)$_{2.2}$Sr$_2$CaCu$_{2}$O$_{8+\delta}$ were
grown by the top solution growth technique.~\cite{Kul,Kul1} 
As-grown samples have plate like shape with the main surface 
coincided with the $\mathbf{ab}$ crystallographic plane.
To provide a flat surface that is always necessary for MO studies
the samples were chemically polished in ethylendiaminetetraacetic 
acid. The final thickness of the samples was  $70\div 100~\mu$m.  
The inductive coil measurements showed $T_c\approx
91$~K with the transition width about 1~K.  
For the sake of easier understanding and comparison we present below 
the images for one typical sample of trapezium shape, shown in Fig.~1. 
Other crystals of different shape exhibited similar results.

   %%%%%%%%%%%%%%%%Fig.1%%%%%%%%%%%%%%%%
\begin{figure}
\includegraphics [width=40mm]{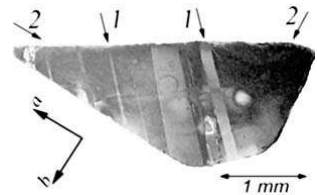}

\caption{\label{fig1}
Polarized light microscope image of the
sample; the crystallographic directions are shown in the left
bottom corner. The directions of twin boundaries and laminar
structure are shown by arrows 1 and 2, respectively. The laminar
structure is invisible in optics.}
\end{figure}

The $\theta-2\theta$ X-ray scanning revealed typical for Bi2212:Pb 
planar defects structure, the twins and laminae.~\cite{Kul,Kul1}
The twin boundaries are parallel to the $\mathbf{c}$-axis and 
coincide with the bisectrix of the angle between $\mathbf{a}$- 
and $\mathbf{b}$-axis. Twins are seen in the presented polarized-light 
image as stripes with light and dark contrast, some of them are marked 
by arrows 1 in Fig.~1. The laminar structure coincides with 
$\mathbf{ac}$-plane. It is parallel with two trapezium sides in Fig.~1 
(the directions of  invisible in polarized light laminae are shown 
by arrows~2).

\section{Experimental} \label{Exp}

The MO studies were performed in the temperature range from 12~K
to $T_c$. The distribution of the transverse magnetic induction
$\mathbf B_z$ was observed by means of standard MO
technique.~\cite{revMO,revMO1,MO} The indicator films used in the study 
allow us to correctly reconstruct the magnetic induction distribution in
$\mathbf H_z$ fields up to 2000 Oe. The MO images were taken by the  EDC1000
digital video camera of fixed sensitivity and variable exposure. 
The brightness of the images is a function of magnetic induction.  
Taking benefits of constant film sensitivity within the temperature
range $12\div 150$~K we calibrate the brightness with respect to the 
induction value. For this purpose we recorded the MO images 
at a set of values of the $\mathbf H_z$ and $\mathbf H_{ab}$ at $T$ slightly above
 $T_c$. As a result, the field mapping and profiles of the transverse 
 magnetic induction along and across the applied in-plane magnetic 
 field were obtained. 

The transverse magnetic field $\mathbf H_z$ parallel to the crystal 
$\mathbf{c}$-axis was generated by a solenoidal coil and varies from 
0 to $\pm 1200$~Oe. The in-plane magnetic
field $\mathbf{H}_{ab}$ was produced by Helmholtz
coils with a soft magnetic core. The uniformity of the field
better than 1~\% was attained across the sample.
We were capable to rotate the field $\mathbf{H}_{ab}$ in any
direction and to change its value from 0 to 1800~Oe. The
orientation of the in-plane field was controlled by the MO
technique with an accuracy about 10$^{-3}$ rad.
The experiments were performed on samples cooled from room 
temperature either with or without $\mathbf H_{ab}$ 
(FC or ZFC regime, respectively).

\section{Magnetooptic observations of flux penetration} \label{MO}

\subsection{Zero in-plane field}

The transverse magnetic flux penetrates into the sample under the 
growing $\mathbf H_{z}$ through a few "weak points", which are 
located at the positions where the twin boundaries intersect the 
sample edges, compare Figs.~1 and~2. The flux enters through the same 
points within the temperature range $12\div 54$~K.
The penetrated flux looks like a bubbles attached to the sample 
edges, Figs.~2a-2c, White spots near the left sample edge in Fig.~2
correspond to the entered magnetic flux; the brighter is the spot 
the higher is the induction. 
    
	%%%%%%%%%%%%%%%%Fig.2%%%%%%%%%%%%%%%%
\begin{figure}
\includegraphics [width=80mm]{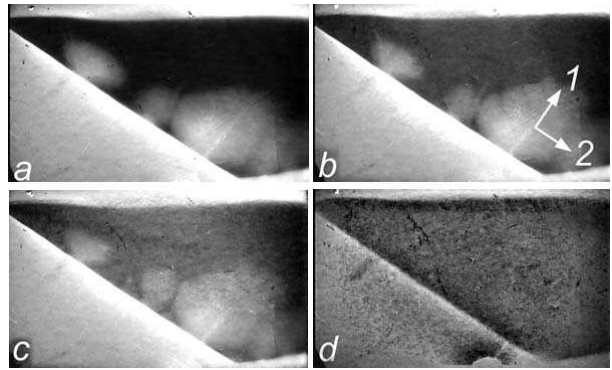}
\caption{\label{fig2} MO images of the transverse magnetic
field penetration at $H_{ab}=$0; (a) $T=24$~K, $H_z=149$~Oe, (b)
$T=37$~K, $H_z=75$~Oe, (c) $T=51$~K, $H_z=46$~Oe and (d)$T=54.5$~K,
$H_z=34$~Oe; white arrows 1 and 2 indicate the directions across
and along the laminar structure.}
\end{figure}

 An increase of $\mathbf H_{z}$ expands flux 'bubbles'
living then attached to the edge. 
The same penetration depth is reached at lower fields with  
temperature increase. The shape of the entered flux is near the 
same at all temperatures below $54$~K.

The flux localization allows us to  easily determine 
anisotropy of the flux penetration in the $\mathbf{ab}$-plane based on 
the bubble's shape. This anisotropy 
is remarkably smaller than expected from the literature 
data about critical current anisotropy.~\cite{Pb4} 
Note, the twin structure in our sample is arranged so that 
the $\mathbf {a}$-axis is 
parallel to the left edge of the sample in Fig.~1. 
Therefore the current along the edge should be $(1\div 4)$-times higher 
than the current across the edge.~\cite{Pb4} Hence, the flux penetration
across the edge should be shorter than along the edge.
Moreover, the higher is the temperature the greater should be anisotropy.
In our experiments most of the observed flux 'bubbles' show rather small
anisotropy at all temperatures below $54$~K. Only flux spot located 
very near to the sharp corner of the sample has a pronounced flux 
penetration anisotropy, Fig.~2.
However this anisotropy could be explained by an influence  of laminar 
structure as well as by Meissner current configuration   
along the nearest sample edges. 

Evidently, the flux penetration characterizes the current anisotropy 
only qualitatively because the penetration is determined by  
distribution of all screening currents in the sample.         

The current distribution 
can be calculated from magnetic induction map.~\cite{MO4,Pb4} 
However the absolute current value can be obtained only under some 
hypothetical approximations of space current distribution. 
Therefore the derivative of the induction taken in the direction 
perpendicular to the flux front, $\partial \mathbf B_z(\alpha)/\partial \mathbf r$, 
which is proportional to the current value, is given in the figures below.
The profiles measured along ($\alpha=90^0$) and across ($\alpha=0^0$) 
laminar structure are given in Fig.~3 
(corresponding directions are marked by arrows in Fig.~2b).
%%%%%%%%%%%%%%%%Fig.3%%%%%%%%%%%%%%%%
\begin{figure}
\includegraphics [width=50mm]{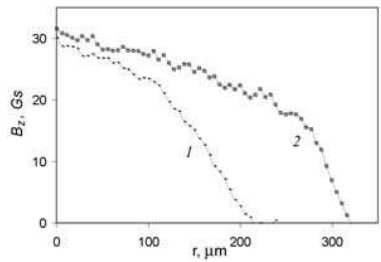}
\caption{\label{fig3} Magnetic flux distribution $B_z(r)$ along
two directions, marked by arrows 1 and 2 in Fig.~2~(a),
$H_z=38$~Oe, $H_{ab}=$0, $T=30$~K. The coordinate origin $r=$0 is
chosen at the point with the maximum induction magnitude.}
\end{figure}

The magnetic induction in the "bubble" is a slowly varying function 
of coordinates near the center of the flux and decays steeply at 
the periphery. This steep part is used to determine 
$\partial \mathbf B_z/\partial \mathbf r$ proportional to the current.
The magnetic induction slopes demonstrate definite anisotropy, 
which could be attributed to some extent to the flux pinning 
by the laminar structure.
   %%%%%%%%%%%%%%%%Fig.4%%%%%%%%%%%%%%%%
\begin{figure}
\includegraphics [width=50mm]{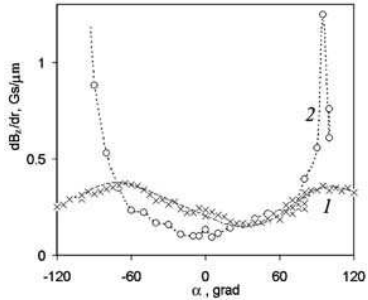}
\caption{\label{fig4} Angular dependence of $|\partial
B_z/\partial r|$ at $T=30$~K and $H_z=38$~Oe; curves 1 and 2 are
obtained at $H_{ab}=0$ and 1800~Oe, respectively. Zero angle
corresponds to the direction indicated by arrow 1 in Fig.~2(b).}
\end{figure}
 
The slope of $\mathbf B_z(\mathbf r)$ varies with the direction, 
Fig.~4 (curve 1), that is determined as an angle, $\alpha$, 
between the directions of vector $\mathbf r$ and the perpendicular 
to the laminae, marked by 1 in Fig.~2b.
It should be noted that the higher is the 
$\partial \mathbf B_z/\partial \mathbf r$ 
in some direction, the higher is the current $\mathbf J(\alpha)$, 
that flow in perpendicular direction.
We find that the current anisotropy does not follow exactly the 
expected anisotropy with minimum current along the 
laminae, $\mathbf J_{min}\neq \mathbf J(0)$  and maximum current 
across them, $\mathbf J_{max}\neq \mathbf J(90)$. We find shift 
of both directions 
at which the current reaches the extremum for $30^0$-clockwise.  
The current anisotropy to compare with the literature data~\cite{Pb4}
could be characterized by the coefficient 
$\mathbf k_J$ defined as the ratio of maximum and minimum derivatives 
taken at the flux sport periphery, 
$k_J = (\partial \mathbf B_z/\partial \mathbf r)_{max}/ 
(\partial \mathbf B_z/\partial \mathbf r)_{min}$.
The $\mathbf k_J$ does not acceed~2 in temperature range from 12~K up 
to 50~K and fields up to 300~Oe. 

The magnetic field penetration behavior changes with temperature,
compare  Figs.~2a-2d. The magnetic flux 'bubbles' remane attached 
to the flux entrance points with field rise while $T < T_m=54 \pm 2$~K. 
However the flux creep grows with the temperature. 
The flux configuration is quasi-stable at $T=12$~K; 
the flux enters only a few percent deeper into the sample in ten 
minute after $\mathbf H_z$ increase at $T=30$~K; the flux spreads in 
seconds for 20\% deeper at $T=51$~K still remaining separated 
"bubbles" with definite $\mathbf B_z$-slopes at flux periphery 
if $\mathbf H_z$ is small enough. 
The shape of the entered flux and the induction distribution profiles 
looks very similar in this temperature range for the flux that just 
enter the sample and for awhile. 

At $T\geq T_m$, the flux penetration behavior changes
drastically. The magnetic flux starts to penetrate the
superconductor at some "weak" points and spreads very fast 
in the whole sample volume.  
Only frame by frame browsing of video-records allows us to 
understand the flux behavior. We found that the flux spreads 
very fast through the geometrical center of the sample, 
changing the shape from the beachcomber
for smooth flux distribution dropped near the sample edges by Meissner 
current. All process runs in a time less than 0.1~s. 
So during this short time the flux distribution becomes   
typical for undoped Bi2212, Fig.~2d, that is determined by Meissner 
current in the absence of pinning.~\cite{In}

\subsection{MO studies in crossed fields}

The in-plane magnetic field, $\mathbf {H}_{ab}$, changes the flux 
penetration pattern if $T<T_m$. 
The magnetic field enters the superconductor from the same weak
points as without $\mathbf {H}_{ab}$. However the flux
diffuses predominantly along the direction of the $\mathbf{H}_{ab}$. 
For this reason, the entered flux looks now rather like stripes, 
extended along the $\mathbf {H}_{ab}$.  

   %%%%%%%%%%%%%%%%Fig.5%%%%%%%%%%%%%%%%
\begin{figure}
\includegraphics [width=80mm]{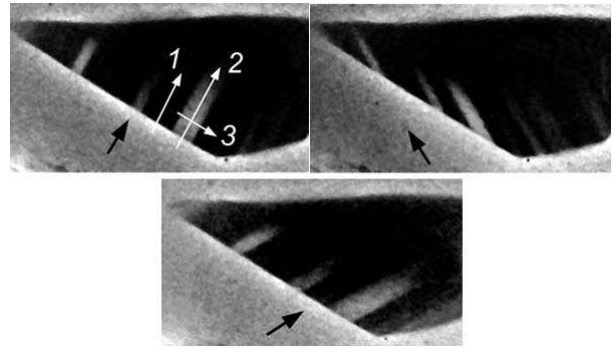}
\caption{\label{fig5} MO images of the flux distribution 
under the action of crossed penetration into
the sample cooled in $H_{ab}=1800$~Oe ($H_z=60$~Oe and $T=30$~K).
Different images correspond to different directions of
$\mathbf{H}_{ab}$ indicated by black arrows in the figures.
White arrows 1, 2 and 3 indicate the chosen direction for $B_z(r)$,
given in Fig.~6}
\end{figure}

Typical pictures of the $\mathbf B_z$ distribution in the presence 
of perpendicular magnetic flux penetration
into the sample cooled in the in-plane magnetic field (FC regime)
are shown in Fig.~5 at fixed value of the transverse field $\mathbf H_z$
and temperature $T<T_m$. Different images in the figure correspond
to different directions of the vector $\mathbf{H}_{ab}$. 
The flux penetration depth and the magnetic induction magnitude rise 
with the increase of transverse field $\mathbf H_z$. 
The anisotropy of magnetic flux penetration increases monotone
with the in-plane magnetic field; the entered flux 'stripes' 
become narrow and longer. The anisotropy  is the same 
in FC and ZFC regimes. 

The appearance of the penetration anisotropy induced by the in-plane 
field in Bi2212:Pb was not reported till now. Such behavior of the 
magnetic flux is observed within the temperature range from 12 to
$54\pm2$~K. This type of the field penetration is analogous to
that usually observed in YBCO single crystals. 

The Fig.~6 shows three profiles of the magnetic field induction 
taken along the directions shown in Fig.~5. 
Curve~1 is the coordinate dependence of $\mathbf B_z$ along the
$\mathbf{H}_{ab}$ direction ($\mathbf x$-axis) measured between two weak
points where the magnetic flux penetration is screened by the
supercurrents. Some growth of the magnetic field near the edge is
due to the non-zero demagnetizing factor of the sample. This field 
'hill' is always observed near thin crystal. Curve~2 is
the magnetic induction profile $\mathbf B_z(\mathbf x)$ scanned 
in the same direction but near the weak point within the band of 
the magnetic flux penetration into the superconductor. This profile 
exhibits a definite drop near the surface due to the Meissner current. 
 
   %%%%%%%%%%%%%%%%Fig.6%%%%%%%%%%%%%%%%
\begin{figure}
\includegraphics [width=50mm]{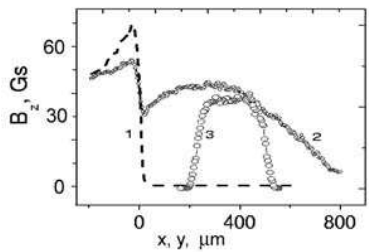}

\caption{\label{fig6} 
Profiles of the magnetic field induction
at $T=30$~K, $H_{ab}=650$~Oe, and $H_z=60$~Oe; curve 1 is the
induction profile along $\mathbf{H_{ab}}$ far from weak points,
curve 2 and 3 are obtained near a weak point along and across the
direction of in-plane field, respectively.}
\end{figure}

In the bulk of the sample, the entering magnetic flux has a variable
slope similar to that in the case of $\mathbf H_{ab}=$0, Fig.~3; namely,
$\mathbf B_z(x)$ varies slowly near the magnetic induction maximum and
drops down almost linearly at the flux periphery. Profile~3 is 
obtained in the same flux penetration zone as profile~2 but in
the direction perpendicular with respect to the vector
$\mathbf{H}_{ab}$ ($\mathbf y$-axis). This profile also consists of parts
with different slopes, that is, a central part with a small flux
gradient and two peripheral parts with the much steeper and almost
linear slopes. The value of the transverse (with respect to the
in-plane field) slope $\partial \mathbf B_z/\partial \mathbf y$ 
is much larger than the longitudinal one 
$\partial \mathbf B_z/\partial \mathbf x$ and both of
them are much smaller than the slope of the magnetic induction due
to the Meissner current, compare profiles in Fig.~6. 

The in-plane field changes the gradient of the magnetic induction, 
$\partial \mathbf B_z/\partial \mathbf r$, along all flux periphery. 
The corresponding angle dependence, 
$\partial \mathbf B_z/\partial \mathbf r (\alpha)$ measured in the 
same manner as in the case of $\mathbf H_{ab}=$0,  is shown in 
Fig.~4 (curve 2). The distribution is obtained for the stripe-like flux
entered the crystal at $T=30$~K following FC under $\mathbf{H}_{ab}=$~1800 Oe. 
The $\mathbf{H}_{ab}$ is directed along laminae. 
The $\partial \mathbf B_z/\partial \mathbf r (\alpha)$ has minimum in 
this namely direction and increases sharply near perpendicular direction. 
It is evident the in-plane field induced anisotropy is much stronger 
than the anisotropy due to the laminar structure. 

The slopes of the induction and the currents proportional to the slopes, 
changes with the value of the in-plane field. The induction 
derivative along the in-plane field decreases with $\mathbf H_{ab}$ 
while the derivative across this direction increases, Fig.~7. 
Hence, $\mathbf J_{sx}(\mathbf H_{ab})$ is an increasing
function while $\mathbf J_{sy}(\mathbf H_{ab})$ is a decreasing one. 
Accordingly, the current anisotropy $\mathbf k_J$ is a rising function of 
$\mathbf H_{ab}$, Fig.~7. So, the current in Bi2212:Pb behaves 
with the in-plane field at $<T<T_m$ in the same manner as in YBCO. 

   %%%%%%%%%%%%%%%%Fig.7%%%%%%%%%%%%%%%%
\begin{figure}
\includegraphics [width=50mm]{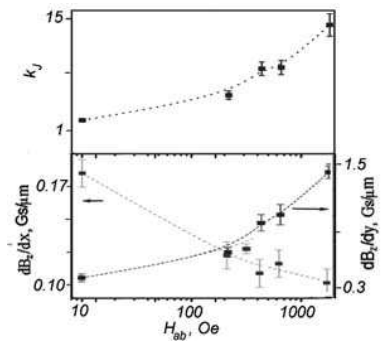}
\caption{\label{fig7} Dependences $|\partial B_z/\partial x|$
and $|\partial B_z/\partial y|$ vs $H_{ab}$ and $k_J$ for $T=30$~K and
$H_z=77$~Oe.}
\end{figure}

The magnetic flux penetration depth grows monotone with $T$ in
the temperature range $12$~K$<T<T_m$, Fig.~8. It is evident the
geometrical anisotropy of flux penetration $\mathbf k$ increases 
with $T$. The analysis of the magnetic field profiles $B_z(r)$ 
reveals that the current anisotropy $k_J$ is a growing function 
of temperature as well despite both screening current components 
$\mathbf J_{sx}$ and $\mathbf J_{sy}$ decrease with temperature, 
Fig.~9. The curves $k(T)$, $k_J(T)$ are also presented in Fig.~9. 
These values increase with temperature approaching to saturation 
at $T\approx 40\div 45$~K for $k(T)$ and
$25\div 30$~K for $k_J(T)$.

   %%%%%%%%%%%%%%%%Fig.8%%%%%%%%%%%%%%%%
\begin{figure}
\includegraphics [width=80mm]{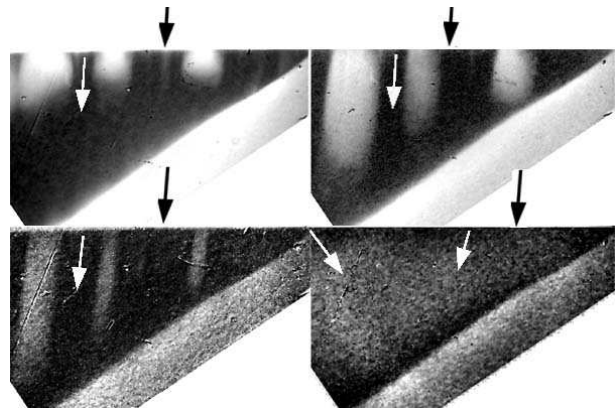}
\caption{\label{fig8} MO images at $H_{ab}=1800$~Oe and
different temperatures; (a) $T=17$~K and $H_z=302$~Oe,  
(b) 29~K and 154~Oe, (c) 43~K and 54~Oe, (d) 56~K and 40~Oe. 
Black arrows show the in-plane magnetic field direction,
white arrows indicate the preferable direction of magnetic 
flux diffusion.}
\end{figure}

The picture of the magnetic field penetration changes dramatically
if temperature exceeds the transition value $T_m$ (see Fig.~8d)
just as in the case of $\mathbf H_{ab}=0$. The flux motion becomes 
independent of the in-plane magnetic field direction. The vortex 
lines enter the sample through a weak points in the same way as at 
lower temperatures, but move to definite geometric 'center'. 
The position of this 'center' is determined by the sample shape 
only and does not depend on the direction and value of the in-plane 
field. The magnetic flux fills the sample volume in a time less than 
0.04~s. Browsing of the MO frame sequence allows us to reveal the 
direction of the magnetic flux motion indicated by white arrows
in Figs.~8, black arrows indicate directions of the in-plane field. 
Such mode of flux penetration in crossed fields is similar to that
observed in Bi2212 single crystals in the absence of pinning. 
The final vortices distribution in our samples at $T>T_m$ is also 
similar to Bi2212 with typical dome-shape determined by the 
geometric barrier. It should be emphasized that the crossover 
temperature $T_m=54\pm 2$~K is independent of the magnetic field 
within the field ranges $\mathbf H_{ab}=0\div 1800$~Oe and
$\mathbf H_{z}=0\div 300$~Oe.

   %%%%%%%%%%%%%%%%Fig.9%%%%%%%%%%%%%%%%
\begin{figure}
\includegraphics [width=0.4\textwidth]{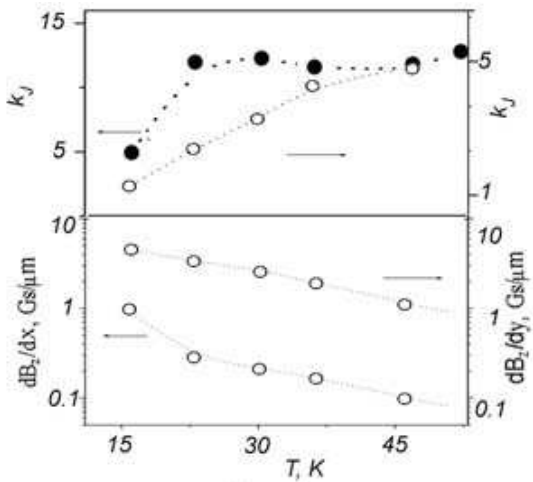}
\caption{\label{fig9} Dependences $|\partial B_z/\partial x|$
and $|\partial B_z/\partial y|$ vs $T$ and $k_J$ under the crossed 
fields $H_{ab}=1800$~Oe and $H_z=77$~Oe. The anisotropy of the flux 
penetration $k(T)$ is shown also}
\end{figure}

We studied also these phenomena in samples with the same Pb doping 
but without twins. The magnetic 
flux enters such samples by a wide pillow-like front. The in-plane 
magnetic field influence the depth of the flux penetration and the
 magnetic induction slope. The anisotropy induced by the in-plane 
 field was observed in these samples only at $T<T_m$. 
 
The described above behavior of the magnetic flux in the FC regime
is similar to the picture of the magnetic field penetration in the
ZFC mode.

\section{Discussion} \label{Disc}

\subsection{Magnetic flux dynamics below $T_m$}

The MO studies in crossed fields reveal that supercurrents of
three different types screen the magnetic flux entering into the
sample. The largest current (see curve~1 in~Fig.~6) flows near the
sample surface far from the 'weak' points. It can be associated with
the Meissner current $\mathbf J_{sm}$. The $\mathbf J_{sm}$ is 
independent of the in-plane magnetic field and is by order of 
magnitude higher than two other currents $\mathbf J_{cx}$ and 
$\mathbf J_{cy}$ flowing in the sample bulk along and across the 
direction of the vector $\mathbf{H}_{ab}$, 
$\mathbf J_{cy}\ll \mathbf J_{cx}\ll \mathbf J_{sm}$. 
Naturally, the values of the screening currents
$\mathbf J_{sm}$, $\mathbf J_{cx}$, and $\mathbf J_{cy}$ decrease 
with the temperature.

We observed $\mathbf J_{cx}\propto |\partial \mathbf B_z/\partial
y|$ grows monotone with $\mathbf H_{ab}$, Fig.~7. Really,   
the current along the in-plane field does not act by the Lorentz
force on the in-plane vortices, as discussed in the Introduction. 
In such force-free configuration the critical current density should 
increase with the increase of $\mathbf H_{ab}$.~\cite{CampIv} 
The growth of $\mathbf J_{cx}$ with $\mathbf H_{ab}$ should take place  
for anisotropic Abrikosov-like flux line structures as well as for
3D correlated stacks of pancake-like vortices. 

The current across the in-plane field, 
$\mathbf J_{cy}\propto |\partial \mathbf B_z/\partial x|$, 
decreases with $\mathbf H_{ab}$ as one often observe for pinning 
controlled current, Fig.~9.

The penetration and current anisotropy increases with temperature while 
$T<T_m$, Fig.~12. Such anisotropy growth was found for YBCO single
crystals.~\cite{In} This could be connected with the rise of the coherence 
lengths with~$T$. The superconducting correlations
between different CuO planes become stronger giving rise to a stronger 
interaction between the in-plane and transverse magnetic flux. 

The presented results show that the magnetic properties of our samples 
at $T<T_m$ are in many features analogous to that of YBCO. This allows 
us to assume the existence of rather strong 3D correlations in the vortex 
system of Bi2212:Pb. 

\subsection{FLL phase above $T_m$ and transition temperature}

The apparent picture of the magnetic flux penetration changes drastically 
at $T=T_m$. The observed disappearance of the intercoupling
between the in-plane and transverse magnetization could find the 
reasonable explanation in terms of a decay of correlations between 
pancakes located in different CuO planes.~\cite{In,kes,Letter} 
A possible alternative explanation of the crossover in magnetization 
behavior at $T=T_m$ is vortex depinning that means the $T_m(B)$ should 
be treated as a irreversibility line. Really, the flux creep rate rises 
and the pinning diminishes with $T$  approaching to $T_m$. 
However, at this temperature 
the correlation between the direction of the entering transverse magnetic 
flux and the in-plane field decays in a step-like manner also. The latter 
effect could not be explained in terms of the thermal depinning. Moreover, 
the found crossover temperature $T_m$ is independent of the magnetic field, 
while the magnetic field dependence of the irreversibility line is usually 
strong for Bi-based systems.~\cite{ir1,ir2,ir3,ir4} Thus, an attempt to 
attribute thermal depinning fails. In the same time, a decay of a 
3D-correlations in the flux line system should be supplemented by the 
reducing of the bulk pinning and by disappearance of correlations 
between the transverse and in-plane flux. Therefore, we consider  
the 3D-2D crossover as a more realistic explanation of the observed 
transition at $T=T_m$. 

Such type of 3D-2D transition could occur due to a melting of the
vortex structure during which the strong correlated stacks of pancakes 
melt in disordered gas or liquid of 2D vortices.~\cite{Fein,Blat,Brandt} 
An indication that the observed change in FLL properties at $T_m$ is due 
to some first-order phase transition is the behavior of the magnetic 
flux penetration anisotropy versus the temperature. The anisotropy 
increases with $T$ and reduces abruptly at $T=T_m$. The transition of 
the pancake stacks into the phase of non-correlated 2D vortices should 
lead to a considerable increase of the thermal creep, e.g. due to 
diminishing of the activation volume and due to corresponding decrease 
of the effective pinning,~\cite{Blat} that is observed in the experiment. 

Following common conceptions,~\cite{Fein} the melting temperature
$T_m$ can be estimated by equating the characteristic energies of
the FLL elastic strain and thermal fluctuations $k_B T$, where
$k_B$ is Boltzmann's constant. The corresponding relation is given
by
\begin{equation}\label{Tm}
k_B T_m = a_L C_{66}a_0^2 d_c,
\end{equation}
where $a_L\ll$1 is Lindeman's constant, $C_{66}$ is the FLL shear
modulus, $a_0$ is the FLL constant, and $d_c$ is an effective
correlation length between the pancakes along 
$\mathbf{c}$-axis.~\cite{Lind1,Lind2} 
The $d_c$ coincides with the distance between neighboring CuO
planes by order of magnitude in strongly anisotropic 
system.~\cite{Fein} The melting temperature $T_m$
defined by Eq.~(\ref{Tm}) is independent of the perpendicular 
magnetic field $B_z$ since $a_0\propto 1/\sqrt{B_z}$
and $C_{66}\propto B_z$.~\cite{Fein,Blat,Brandt} 
The last fact is in agreement with
the results of the present experiments. In the dislocation 
melt approach Lindeman's constant can be estimated as 
$a_L=1/4\pi$.~\cite{Lind1} 
Substituting $C_{66}=B_z\phi_0/(4\pi\lambda_{ab})^2$ and
${a_0}^2=2\phi_0/\sqrt{3}B$ into Eq.~(\ref{Tm}) one finds the
equation for the melting temperature~\cite{Fein,Lind2}
\begin{equation}\label{Tm1}
k_B T_m = \phi_0^2 d_c/32\sqrt{3}\pi^2\lambda_{ab}^2(T_m),
\end{equation}
where $\phi_0$ is the magnetic flux quantum and $\lambda_{ab}(T)$
is the London penetration depth in the $\mathbf{ab}$-plane. 
We find $T_m\approx 50$~K if $d_c = 1-2$~nm and 
$\lambda_{ab}(0)=200-300$~nm, which seems reasonable 
for Bi2212:Pb.~\cite{Pb,Kul} This estimation is not a strong proof 
but some evidence in favor of the 3D-2D transition.  

In disordered 2D phase the correlation between the motion 
of the transverse flux and the in-plane magnetic field is 
significantly lower compared with the 3D correlated system. 
The reduction of the activation volume makes easier the
thermoactivated motion of the non-correlated 2D vortices in any
direction. 

\section {Conclusion} \label{Conc}

The MO studies of Bi2212:Pb single crystals in
crossed magnetic fields revealed that a transition occurs in the
magnetic flux behavior at $T=T_m=54\pm2$~K. The transverse 
magnetic flux at $T<T_m$ behaves like in YBCO spreading 
preferably along the in-plane magnetic field. At $T>T_m$ 
the transverse flux penetrates independent of the in-plane 
magnetic field as in Bi2212 system. 
The anisotropy of the flux penetration increases
with the in-plane magnetic field and temperature at $T<T_m$. 
The transition temperature is independent of the magnetic field. 
The obtained experimental results could be understood within 
the concept of the flux line melting giving rise to the 
transition of 3D correlated stacks of pancakes into disordered 
phase of 2D ones. We believe that the existence of strong 3D 
correlations in the flux line structure due to Pb doping 
is the main reason for enhanced critical current in Bi2212:Pb.

\begin{acknowledgments}

The authors acknowledge M.~V.~Indenbom, V.~V.~Riazanov,
L.~M.~Fisher, I.~F.~Voloshin, A.~V.~Kalinov, I.~K.~Bdikin for
useful discussions, and Alexander von Humboldt Foundation for
financial support for part of the experimental equipment.

This work is supported by INTAS (grant 01--2282), RFBR (grants
02--02--17062 and 00--02--18032) and Russian State Program on
Superconductivity (project 40.012.1.1.11.46).

\end{acknowledgments}

\end{document}